# Beyond ΛCDM: Bits join the Dark Side

Michael Paul Gough

Department of Engineering and Design, University of Sussex,

Brighton, BN1 9QT, UK.

E-Mail: m.p.gough@sussex.ac.uk

**Abstract:** We are encouraged to look beyond ΛCDM as there are no satisfactory explanations for dark energy or dark matter. An evidence based approach applies two foundational principles of information theory to show dark energy and dark matter may be Holographic Dark Information Energy, HDIE. HDIE explains many effects attributed separately to Λ and CDM. HDIE mimics Λ with sufficient energy and an equation of state parameter, $w$= -1.03±0.05 at redshifts $z$<1.35 to account for accelerating expansion. HDIE is clumped around galaxies at densities that distort space-time, explaining many CDM attributed effects. The present ratio of HDIE:baryons ~2.15, required for observed expansion history, is equivalent to a dark matter fraction ~68%, consistent with several galaxy surveys. The HDIE/baryon model is based largely on proven physics, provides a common explanation for dark energy and dark matter, and solves the cosmological coincidence problem. At earlier times, $z$ > ~1.35, HDIE was phantom, $w$ = -1.82±0.08, enabling the model to be falsified. HDIE fits *Planck* dark energy $w_o$-$w_a$ plots at least as well as Λ, and is consistent with other results that suggest dark energy was phantom at earlier times. A new $w$-parameterisation is proposed, as the usual CPL parameterisation is unsuitable for distinguishing between HDIE/baryon and ΛCDM models.

**Keywords:** cosmology, dark energy, dark matter, cosmological constant problem.

## 1. Introduction

ESA *Planck* results [1-3] are compatible with a flat universe containing 68.3% dark energy, DE, 26.8 % dark matter, DM, and only 4.9% ordinary baryonic matter. In the popular "concordance", or standard model, ΛCDM, the dark components DE and DM are provided by Einstein's cosmological constant, Λ, and cold dark matter, CDM, respectively.

The observed accelerating expansion of the universe [4,5] is explained by the action of a DE that exhibits a near constant overall energy density, at least in recent times, and thus has been naturally associated with the cosmological constant, Λ, the energy of a vacuum [6-10]. Theoretical attempts to account for the required DE energy density are out by very many orders of magnitude, and even a zero valued cosmological constant is considered more likely than the observed DE value [8]. The cosmological constant explanation leads us to assume a spatially constant energy density throughout the universe. However, we only know from universe acceleration that total DE increased in near proportion to universe volume over recent times. We do not know how DE is distributed.

Observed galaxy spin profiles and gravitational lensing effects require significantly stronger gravitational effects than can be provided by visible baryon matter alone. These effects have been naturally attributed to DM, an as yet unidentified species of particles that are difficult to observe as they experience negligible interaction with ordinary matter or electromagnetic radiation. Various explanations for DM have considered a range of possible particle types [11] including: weakly interacting massive particles (WIMPS), neutralinos, asymmetric dark matter, MACHOS, axions, and others. Many techniques are being applied with continually improving sensitivity to search for DM [12], ranging from attempts to directly detect DM particles with large volume detectors in quiet locations, attempts to create DM particles inside high energy particle colliders, and attempts to detect radiation generated when DM particles interact with each other, or annihilate. A small residual excess of gamma rays from the galactic centre [13], initially thought to result from DM annihilation, is now considered to be explained by a population of neutron stars or pulsars [14,15]. Extra-galactic gamma rays studied with 6 years of data from the Fermi Large Area Telescope [16] have shown no evidence for DM. The most sensitive WIMP detectors to date, Xenon100, LUX, PandaX-II, and PICO-60 have all failed to detect any DM particles [17-20]. Although the sensitivity of DM detectors continues to improve, there is now less confidence that WIMPS, the favoured DM candidate, will ever be found [21][22].

DE and DM represent the core of the $\Lambda$CDM concordance model, at a combined 95% of the universe, but satisfactory explanations for both phenomena continue to elude us [23,24]. We are therefore encouraged to consider other possibilities, however radical they may at first appear. Accordingly, in this work we ask the question "do we need dark matter?". We suggest that DE may not be evenly distributed but clumped around the structures of ordinary matter. All of the experimental evidence for DM is based solely on its gravitational effects, via gravitational distortions of space-time [12]. Clumps of DE could also produce DM like effects, since such significant concentrations of energy will distort space-time in exactly the same way as an energy equivalent quantity of matter. In this way we may consider a universe consisting primarily of DE and baryons, without the need to invoke the existence of any exotic DM particle species.

Previously it was assumed that the negative pressure aspect of increasing total DE with increasing volume on the universe scale would not allow DE to become clumped. The form of DE proposed here will be clumped and gravitationally attractive on the local scale of galaxies and galaxy clusters, but will exert an overall repulsive effect in the universe. Note that this approach differs from MOND [25] and Dark Fluid [26] theories in that it does not require any extensions or modifications to gravity theory, nor does it introduce new physics.

One model already shown capable of quantitatively explaining DE is holographic dark information energy, HDIE [27-29], essentially the energy equivalence of the information, or 'entropic energy', carried by the universe's baryons. Every baryon effectively carries bits of information [30] and each bit of information has an energy equivalence [31-36]. In recent times, redshifts $z<\sim1$, HDIE exhibits a near constant overall energy density, corresponding to an equation of state parameter, $w\sim-1$, and compatible with the *Planck* 2013 data release [1,29]. Such a source of DE is naturally concentrated around the high temperature structures of ordinary matter: stellar heated gas and dust, stars, galaxies, and galaxy clusters. HDIE has been shown to be strong enough to explain DE and, as it is clumped around structures, these locally enhanced energy densities must add significant additional gravitational distortions to space-time in the vicinity of those structures.

The aim of this paper is to show that an HDIE/baryon model might account for all of the observations previously attributed separately to $\Lambda$ and CDM. The HDIE/baryon model emphasizes a data-centred phenomenological approach to understanding DE and DM, an approach primarily guided by empirical evidence rather than by theory. This work is driven by the many potential attractions of the HDIE/baryon model. Besides removing the need to discover new exotic dark matter particles, we find that this model naturally removes the "why now?" cosmological coincidence problem. Moreover, it enables both unexplained aspects of the ΛCDM model to be replaced with just one phenomenon that has a present energy density quantitatively explained by proven physics.

In sections 2.1 and 2.2 we summarize and update previous HDIE work [29], adding the latest stellar mass density data, to reaffirm that HDIE can explain DE. Section 2.3 extends this DE explanation to show how HDIE might also explain effects previously attributed to DM. Section 2.4 summarises the potential advantages of the HDIE/baryon model over the ΛCDM model. Section 3 then considers the problem of experimentally distinguishing between the two models, identifies limitations of existing data and data parameterisations, and concludes with the suggestion of a simple preferred parameterisation to achieve this distinction.

## 2. Proposed HDIE/baryon model

**2.1 Information energy in the universe.**

Landauer [31,32] argued that information is physical with a minimum energy equivalence of $k_B T ln2$ per bit, where $k_B$ is Boltzmann's constant and $T$ is temperature. Recent laboratory experiments [37-39] have indeed confirmed Landauer's principle by clearly demonstrating this minimum quantity of heat is dissipated when information is erased. Landauer's principle is, in effect, just an expression of the second law of thermodynamics, since information and entropy are identical for the same degrees of freedom, with the only difference in measurement units (1bit=$ln2$ nats). Note entropy or information carried by nature is a scalar field.

| | | Information, $N$ bits | Temperature $T$ °K | Information Energy $N k_B T ln2$, Joules |
|---|---|---|---|---|
| Relics of Big Bang | CMB photons | $10^{88} - 2 \times 10^{89}$ | 2.7 | $3 \times 10^{65} - 6 \times 10^{66}$ |
| | Relic neutrinos | $10^{88} - 5 \times 10^{89}$ | 2 | $2 \times 10^{65} - 10^{67}$ |
| | Relic gravitons | $10^{86} - 6 \times 10^{87}$ | ~1? | $10^{63} - 6 \times 10^{64}$ |
| Dark matter | Cold dark matter | $\sim 2 \times 10^{88}$ | $<10^2$ ? | $< 10^{67}$ |
| Star formation | $10^{22}$ stars | $10^{79} - 10^{81}$ | $\sim 10^7$ | $10^{63} - 10^{65}$ |
| | Stellar heated gas and dust | $\sim 10^{86}$ | $\sim 10^6 - 10^8$ | $\sim 10^{69} - 10^{71}$ |
| Black Holes | Stellar sized BH | $10^{97} - 6 \times 10^{97}$ | $\sim 10^{-7}$ | $10^{67} - 6 \times 10^{67}$ |
| | Super massive BH | $10^{102} - 3 \times 10^{104}$ | $\sim 10^{-14}$ | $10^{65} - 3 \times 10^{67}$ |
| Universe | Holographic bound | $\sim 10^{124}$ | - | - |

**Table 1. Universe information content, temperature, and information energy contributions.**

Table 1. lists the relevant information components of the universe, together with estimates of the quantity of information, $N$, associated with each [27-29,40,41], representative temperatures, $T$, and their resulting information energy, $N k_B T ln2$ contributions.

We see from Table 1 that stellar heated gas and dust, at $10^{69}$-$10^{71}$J, makes the largest information energy contribution to the universe. Although the values of Table 1. are only order of magnitude estimates, we can expect the information energy of stellar heated gas and dust to play a significant role in the universe, since this quantity of energy is of a comparable magnitude to the $\sim 10^{70}$J energy equivalence of the universe's $\sim 10^{53}$kg of ordinary matter.

**2.2 HDIE as Dark Energy.**

As the universe expanded the energy density of matter fell as $a^{-3}$, where $a$ is the universe scale factor ($a$=1 today and related to redshift, $z$, by $a$=1/(1+$z$) ). In order to explain the observed change from decelerating expansion to accelerating expansion in the second half of the universe's history [1-3], we require a DE of near constant average energy density, with an energy today $\sim$2.15 times the matter energy. Within the uncertainties of the estimates of Table 1. we can see that the information energy contribution from stellar heated gas and dust is roughly of the order of magnitude that could account for DE today. Then, we need to show that, in recent times, HDIE also possesses a near constant average energy density with an equation of state parameter, $w \sim$-1, or a total DE energy that increased as $\sim a^3$.

The information energy of stellar heated gas and dust varies over time dependant on just two parameters: information content $N$, and average temperature $T$. Here we assume that $T$ will be governed by the extent of star formation, or the fraction of baryons that are now in stars, while $N$ for any given volume of space is set by a generalised Holographic principle [42-44] as proportional to the bounding area of that space, $N \propto a^2$. The Holographic principle is well established for black holes at the holographic bound [45], but is also considered [44] to apply to any region of space, with recent CMB data analysis emphasizing the evidence for a holographic universe [46], even though the universe is many orders of magnitude below the holographic bound (see Table 1.).

In figure 1 we plot a survey of measured stellar mass densities per co-moving volume as a function of scale size, $a$. The filled symbols [47-63] correspond to data compiled for a recent survey of stellar formation measurements (Table 2 of [64]). A subset of these data was already included in previous HDIE work [27-29], and open symbols [65-76] correspond to those measurements used in that previous HDIE work but not included in this recent survey.

There is a clear growth in star formation with approximately one half of today's stars formed before redshift, $z$=1.3, and 25% formed after $z$=0.7. Despite considerable scatter in measured values there appears to be a significant change around redshift, $z \sim$1.35 from a steep gradient in the past to a weaker gradient in recent times. Fitting straight line power laws (red lines, in Fig.1.) to data points either side of $z$=1.35, we find power law fits of $a^{+1.08\pm0.16}$ for $z$<1.35, and $a^{+3.46\pm0.23}$, for $z$>1.35.

Then we can assume average baryon temperature, $T$, is proportional to the fraction of baryons in stars and thus also varied as $a^{+1.08\pm0.16}$ for $z$<1.35. Thus the total stellar heated gas and dust information energy ($\propto NT$) varied as $a^{+3.08\pm0.16}$, corresponding to near constant energy density, or an equation of state parameter, value $w$=-1.03$\pm$0.05. In comparison, total information energy in the earlier period, $z$>1.35, varied as $a^{+5.46\pm0.23}$, corresponding to a phantom energy with $w$= -1.82$\pm$0.08.

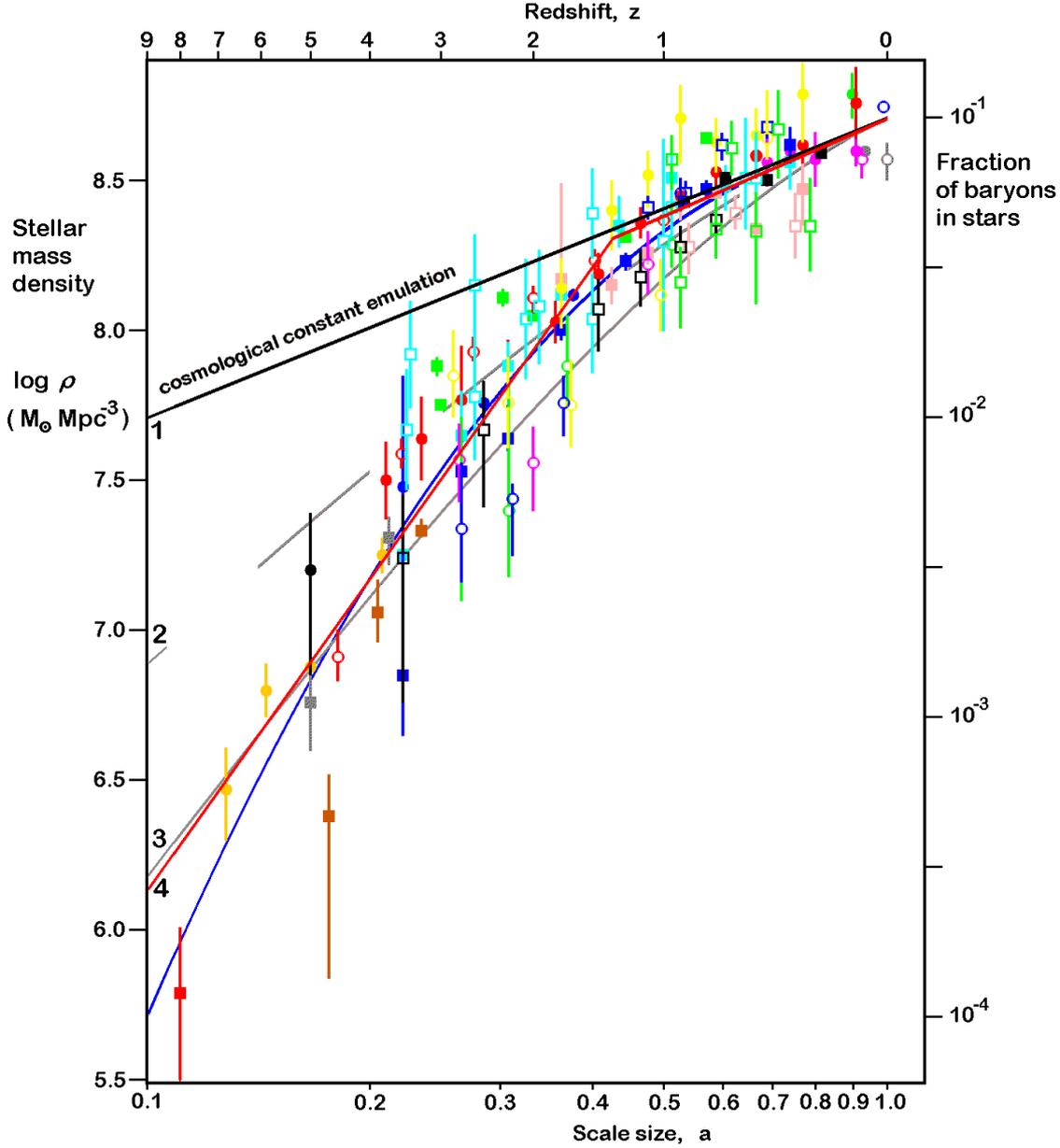

**Fig.1**. Review of stellar mass density measurements for co-moving volumes as a function of universe scale size, *a*.

*Plotted lines*: two red straight lines, power law fits $a^{+1.08\pm0.16}$, ($w=-1.03\pm0.05$) for $z<1.35$, $a^{+3.46\pm0.23}$, ($w=-1.82\pm0.08$) for $z>1.35$; black line, the variation that would be required for HDIE to fully emulate a cosmological constant over all scale sizes; blue curve, simple polynomial fit to all data; and grey curves, variation required to provide HDIE with a CPL-like variation, $w=w_o+(1-a)w_a$ where grey continuous line $w_o=-1.0$ and $w_a=-0.8$, and grey dashed line $w_o=-1.0$ and $w_a=-0.45$. *Numbered lines*, 1 to 4 correspond to parameterisation cases considered in section 3.3.

*Source references. Filled symbols:* grey circle[47]; dark green circle[48]; magenta circle[49]; pink square[50]; red circle[51]; cyan square[52]; blue square[53]; yellow circle[54]; black square[55]; green square[56]; blue circle[57]; dark green square[58]; brown square[59]; orange circle[60]; grey square[61]; black circle[62]; red square[63]. *Open symbols:* grey circle[65]; dark green circle[66]; magenta circle[67]; pink square[68]; red circle[69]; cyan square[70]; blue square[71]; yellow circle[72]; black square[73]; green square[74]; blue circle[75]; dark green square[76].

We see that stellar heated gas and dust in the recent period, $z<1.35$, could then explain DE, since $T$ closely follows the $a^{+1}$ gradient that would lead to a total HDIE varying as $a^{+3}$, to effectively emulate a cosmological constant ($w=-1$, black line in Fig.1). HDIE can therefore explain DE quantitatively, accounting for both the present energy density value and the recent period of near constant overall energy density.

It is worth noting two further consequences of an HDIE explanation for DE.

Star formation had to have advanced sufficiently before HDIE was strong enough to affect universe expansion. Star formation had also to have advanced sufficiently for the likelihood of intelligent beings evolving to observe an accelerating expanding universe. Then HDIE effectively removes the "why now?" cosmological coincidence problem.

The advent of accelerating expansion has been associated [77,78] with causing a general reduction in galaxy merging, and structure formation, evident in Fig.1 as the stellar mass density gradient change from $a^{+3.46\pm0.23}$ to $a^{+1.08\pm0.16}$. Once HDIE was strong enough to initiate acceleration, this in turn inhibited star formation and consequently limited HDIE itself. We expect this feedback mechanism to naturally cap the star formation gradient around a stable value $\sim a^{+1}$, constraining energy density to a constant value around the observed DE energy density. Feedback from HDIE could therefore explain the timing of the change in stellar formation rate around $z\sim1.35$, the power law value after $z\sim1.35$, and the present ratio of DE energy to matter energy.

## 2.3 HDIE imitates Dark Matter.

All of the effects attributed to both DE and DM only occur through the action of gravitational forces. There is no evidence of DE or DM interacting with ordinary baryonic matter or photons through any of the other fundamental forces. This similarity argues for a common explanation.

HDIE is naturally located where baryons occur at high temperature and density and where HDIE energy densities should be sufficient to add significantly to baryon distortions of space-time. As there is no separate DM component in the HDIE/baryon model, we expect that the location of hot baryons will fully specify where the DM attributed effects will occur. Indeed, a high correlation has been found [79, 80] between the observed galaxy radial acceleration and that predicted from baryons, based on a total of 240 galaxies of various morphologies: 153 late-type galaxies, 25 early-type galaxies and 62 dwarf spheroidals. There is very little scatter and this strong empirical relation shows all galaxies studied follow the same radial acceleration relation, showing the dark matter contribution to be fully specified by the baryons. Thus dark and baryonic masses exhibit a strong coupling that is difficult for the ΛCDM model to explain, but would follow directly from the HDIE/baryon model (and possibly also from the MOND model).

Observations of clusters of galaxies [81] show that the brightest galaxies are almost always found in the middle of those locations where gravitational lensing indicates the DM contribution is maximum. Clearly, this property is also consistent with an HDIE explanation since HDIE is proportional to temperature. HDIE also fits with the favoured bottom up hierarchical structure formation with smaller objects forming first and effectively promoting the formation of larger structures, resembling CDM rather than hot DM. In the ΛCDM model gravitational lensing effects are due to higher densities of DM which have led to increased structure formation and brighter galaxies at those locations. In the HDIE/baryon model as galaxies increase in brightness with increasing temperatures, higher entropies, and thus higher HDIE densities lead to both the observed gravitational lensing DM-like effects and further increases in structure formation.

The present universe wide fraction $f_{HDIE}$~68% required for HDIE to explain the universe expansion history is then equivalent to an average DM fraction, $f_{DM}$~68%, a lower value than the $f_{DM}$~85% of the *Planck* ΛCDM model [1-3]. It is beyond the scope of this work to survey all measured $f_{DM}$ values of astrophysical objects, ranging from globular clusters containing little DM to dwarf galaxies dominated by DM. Two random examples illustrate the wide range. One edge-on lensing galaxy was found [82] to have $f_{DM}$= 25%-35%, while one ultra diffuse galaxy [83] has $f_{DM}$>99%.

However, a survey [84] of 1.7 x10$^5$ massive early-type galaxies $z$<0.33 yields $f_{DM}$ = 53%-72% within those galaxies' effective radius (radius that defines the sphere responsible for one half of the galaxy's light emission). Another survey of 584 typical star-forming galaxies, $z$=0.8-1.0 [85] finds $f_{DM}$= 65±12%. Note the present value HDIE $f_{HDIE}$~68% lies in the middle of both survey ranges while ΛCDM $f_{DM}$~85% lies outside.

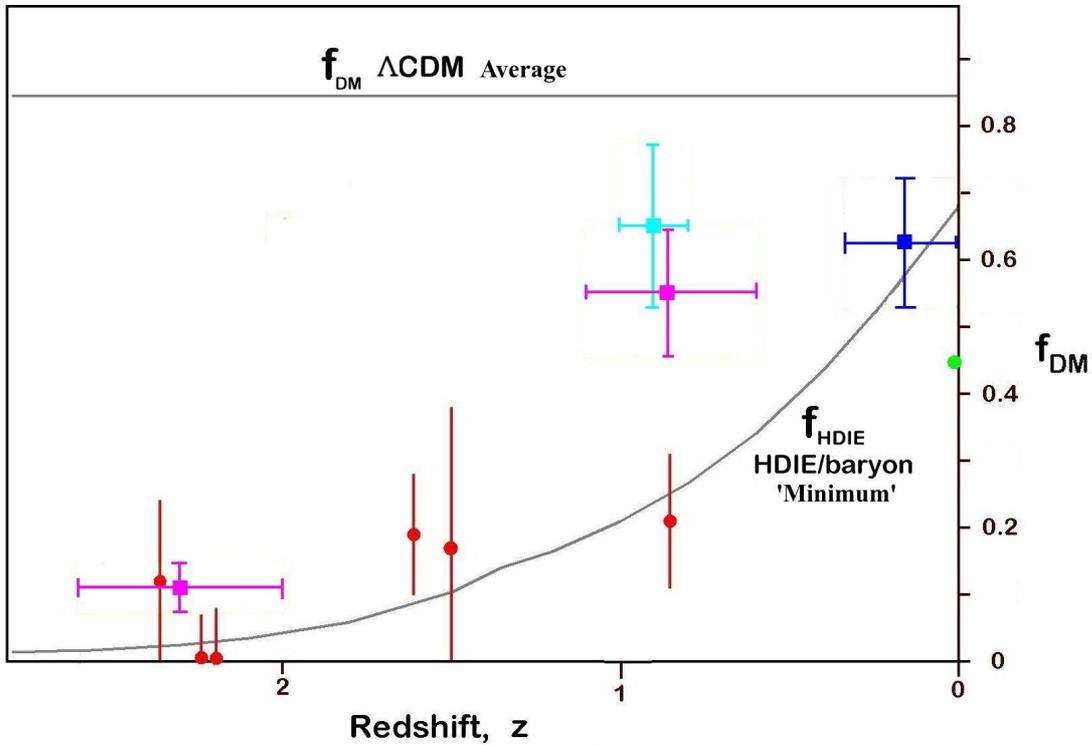

**Fig.2. Observed $f_{DM}$ compared with universe average ΛCDM $f_{DM}$ and HDIE/baryon $f_{HDIE}$.** Survey results, square symbols: Dark Blue, 1.7 x10$^5$ massive early-type galaxies $z$<0.33 [84], Light Blue, 584 typical star-forming galaxies, $z$=0.8-1.0 [85], Purple, 92 star forming galaxies, $z$=2.0-2.6 and 106 star forming galaxies, $z$=0.6-1.1 [86]. Individual galaxies, circle symbols: Red, six early star forming galaxies [86]; Green, Milky Way galaxy [87]. The $f_{HDIE}$ curve is calculated from the power law fits to the star formation data of Fig.1, assuming present $f_{HDIE}$=68%.

While values of $f_{HDIE}$ should be high in the high temperature baryon objects that we can observe, universe average $f_{HDIE}$ would have been much lower at earlier times when only a small fraction of baryons had formed stars. Therefore, we expect observable object $f_{DM}$ values at or above the universe average $f_{HDIE}$ value at that redshift ($f_{HDIE}$ effectively provides a minimum value). In contrast, universe average ΛCDM $f_{DM}$ should have remained constant, independent of redshift, and observed object $f_{DM}$ values should be found distributed approximately evenly about $f_{DM}$ ~85%. Fig.2. compares these universe average values with several galaxy surveys [84-86] and with six early star forming disk galaxies [86]. The variation in DM contribution illustrated in Fig.2 emphasizes the relative absence of DM in early massive star forming galaxies of 10 billion years ago [86]. Note surveys of early star forming galaxies made with the same instruments and analysis techniques, but in two different redshift ranges [86] (purple squares in Fig.2) show that galaxies at $z$=2.0-2.6 clearly have lower $f_{DM}$ values than those at $z$=0.6-1.1. Overall, both survey and individually measured values of $f_{DM}$ are more consistent with the time varying minimum predicted by the HDIE/baryon model than with the fixed 85% average value of the ΛCDM model.

Table 1 shows that there is more HDIE associated with stellar heated gas and dust than with the stars themselves. In addition, X-ray measurements have found the majority of baryons exist as warm/hot gas disconnected from stars: in galaxies as spherical haloes at T~$10^6$ to$10^7$ [88]; in the intracluster medium of galaxy clusters at T~$10^7$ to $10^8$; and in the intergalactic medium at T~$10^5$ to $10^7$. We can therefore expect information energy contributions from other forms of hot gas to contribute to HDIE besides the contribution from stellar heated gas and dust. The combined distribution of HDIE around a galaxy may more likely resemble the galactic halo than the typical disc or spiral of stars within the galaxy, and thus be similar to the distributions of DM around galaxies required to explain rotation measurements.

Clusters of colliding galaxies are considered to provide some of the strongest evidence for the existence of DM. Optical observations show stars pass through largely unhindered while X-ray observations show the galactic gas clouds, containing the majority of baryons, collide, slowing down or even halting. The location of DM is then identified from lensing measurements [89-91]. A study of the Bullet cluster [89], and of a further 72 mergers [90], both major and minor, finds no evidence for DM deceleration, with the dark mass remaining closely co-located with the stars and structure. Thus DM is found to be not concentrated around the baryon centre of gravity in the galactic gas clouds, and an upper limit is placed on any DM self-interaction. Clearly HDIE could equally explain all of these observations if the dominant contribution to HDIE in these cases were from stellar heated gas and dust that generally passes with the stars straight through.

A study of four galaxies colliding in cluster Abell 3827 [91] show similar characteristics, but one of the galaxy's DM appears to be slowed down. That case might be explained as a combination of the HDIE contribution from stellar heated gas and dust with an additional contribution from the intracluster gas, perhaps heated to higher temperatures by shocks.

The spatial distributions of some galaxies and galaxy clusters have been found to exhibit an "assembly bias" [92]. The way in which those galaxies interact with their DM environments appears to be determined not just by their masses but also by their past formation history. This could also be consistent with an HDIE explanation as information/entropy is a result of not just present processes but also the result of the past history of physical processes that operated on baryons.

The milky way, Andromeda, and Centaurus-A galaxies have a number of satellite dwarf galaxies that orbit in the same plane with the majority co-rotating [93,94]. This observation is difficult to reconcile with ΛCDM as DM should be distributed in a sphere around the parent galaxy with satellite galaxies randomly distributed. However, this observation may be consistent with the HDIE/baryon model where DM like effects follow the location of the parent galaxy's hot baryons.

## 2.4 Summary of potential advantages of the HDIE/baryon model.

Then the HDIE/baryon model can be seen to possess several potential advantages over the ΛCDM concordance model. The HDIE/Baryon model can:

- Quantitatively account for the present DE energy density value with proven physics, using experimentally proven Landauer's Principle with realistic universe entropy estimates.

- Explain why DE has an overall near constant energy density in recent times by combining star formation measurements with the Holographic principle. The Holographic principle is generally accepted for black holes at the holographic bound, but remains only a conjecture for universal application.

- Account in general for many effects previously attributed to DM: galaxy spin anomalies; gravitational lensing; lensing of clusters of colliding galaxies; and galaxy 'assembly bias'.

- Account for galaxy radial acceleration with DM effects fully specified by baryon location, and possibly also similarly account for the alignment of satellite galaxies.

- Provide better agreement with measured DM fractions in galaxy surveys.

- Enable an explanation for DM attributed effects without invoking new exotic and practically undetectable particles, and without requiring the new physics required by MOND and dark fluid theories.

- Allow the cosmological constant to take the more likely zero value.

- Solve the cosmic coincidence problem.

- Account quantitatively for the recent change in star formation rate due to DE feedback.

- Reduce a problem of two unexplained phenomena to a single phenomenon.

- Provide an explanation emphasizing 'simplicity' (wielding Occam's razor) and 'naturalness' (relying on mostly proven physics) with a strong dependence on empirical data [24].

- Significantly reduce our ignorance of the universe. The baryon world which we observe, and are ourselves a part of, would now play a more important role in the universe, representing ~32% of the energy total, while the present and past physical processes acting on that component provide the HDIE information energy component that can account for the remaining ~68% of universe energy.

## 3. Measurements to distinguish between HDIE/baryon and ΛCDM models.

If there were to be a confirmed detection of DM particles it would clearly refute our explanation, at least as an explanation for DM attributed effects. However, we cannot use the on-going failure to confirm a DM particle source as positive support for our explanation - absence of evidence is not evidence of absence. Furthermore, in §2.3 we were only able to explain DM attributed effects by HDIE/baryons in the most general qualitative terms.

Given the potential advantages of the HDIE/baryon model listed in §2.4, we must try to satisfy the Popper requirement [95] of predicting the value of a measurement whereby this explanation can be discounted, or falsified. Fortunately, the HDIE/baryon model predicts a significant specific phantom DE behaviour, $w$~-1.8 for $z$>1.35. Thus, we can expect to make more progress by comparing the different universe expansion predictions of the HDIE/baryon and ΛCDM models at higher z values.

Friedman [96] found solutions to the Einstein equations [97] that allowed Weinberg [98] to express in equation 1 the Hubble parameter, $H(a)$ in terms of the Hubble constant $H_0$ (present Hubble parameter value) and dimensionless energy density parameters, $\Omega$, where the present value of each is expressed as a fraction of today's total energy density so that all $\Omega$ terms add up to unity:

$$(H(a)/H_0)^2 = (\Omega_{CDM} + \Omega_b)\, a^{-3} + \Omega_r\, a^{-4} + \Omega_k\, a^{-2} + \Omega_{DE}\, a^{-3(1+w)} \quad (1)$$

Subscripts: *CDM*, cold dark matter; *b*, baryons; *r*, radiation; *k*, curvature; and *DE*, dark energy. Here the $\Omega_{DE}$ term is expressed more generally than the original $\Omega_\Lambda$.

It is usual to assume curvature $\Omega_k$ is zero, and that the radiation term, $\Omega_r$, for some time has been negligible compared to the total of all matter and DE. Further assumptions inherent in the ΛCDM and HDIE/baryon models then lead to two different descriptions of universe expansion:

1) the ΛCDM model assumes DE is the cosmological constant, $w$= -1,

$$(H(a)/H_0)^2 = (\Omega_{CDM} + \Omega_b)\, a^{-3} + \Omega_\Lambda \quad (2)$$

with *Planck* values: $\Omega_{CDM}$ ~ 0.27, $\Omega_b$ ~ 0.05, and $\Omega_\Lambda$ ~ 0.68.

2) the HDIE/baryon model predicts $\Omega_{CDM}$ = 0 and a dynamic DE with a time varying equation of state parameter, $w(a)$,

$$(H(a)/H_0)^2 = \Omega_b\, a^{-3} + \Omega_{HDIE}\, a^{-3(1+w(a))} \quad (3)$$

with accelerating expansion equivalent values: $\Omega_b$ ~ 0.32, $\Omega_{HDIE}$ ~ 0.68, and $w(a>0.43)$ ~ -1.

*Planck* data [1-3] has shown that the universe is flat so the total energy density today must be close to the critical density, $3H_0^2/8\pi G$, or equivalent in energy density to ~6 hydrogen atoms/m$^3$. Then in the ΛCDM model where the ratios of baryons : DM : DE are 5% : 27% : 68%, respectively, the universe average baryon density should be ~0.3 atoms/m$^3$. However, in HDIE/Baryon model ratios of baryons : DM : DE are 32% : 0% : 68%, so the average density should be around ~6x higher at ~2 atoms/m$^3$. The adoption of a higher density is encouraged by recent results: the observable universe has just been found to contain 10 times more galaxies than previously thought [99] and the ESA Gaia spacecraft has shown that even our own galaxy is much bigger than previously thought [100].

At first sight, universe expansion measurements should easily provide a clear distinction between the descriptions of equations (2) and (3). In the HDIE model, DE and hence the universe expansion rate is dependant on structure formation, and, based on the star formation history measurements of Fig.1, we expect the form of $w(a)$ to be such as to provide $w$= ~ -1.0 for $a$ > ~0.43, ( $z$< ~1.35), and $w$= ~ -1.8 for $a$< ~0.43, ( $z$> ~1.35). Thus, for the same present ratio of DE to all matter (~2.15), both models behave identically at low redshifts, with the only difference restricted to $z$>~1.35, where any DE contribution is difficult to measure as it is swamped by the much higher matter energy density at earlier times.

We can illustrate this problem by looking back in time. The present DE contribution of 68% in both models falls to 14% by $z$=1.35. The cosmological constant contribution in the ΛCDM model then continues to fall to 7% at $z$=2, and 1.6% by $z$=4, while in the HDIE/baryon model we expect HDIE to fall more rapidly, down to ~3% at $z$=2, contributing only ~0.25% at $z$=4. Note that the differences between models in total energy at $z$=2 and $z$=4 of ~4% and ~1.35% correspond to differences in Hubble parameter, $H$, of only ~2% and ~0.7%, respectively, requiring very high precision expansion measurements to enable the necessary distinction between models.

The expansion rate variation over time is determined by combining a number of different measurement types. Unfortunately, most of the non-CMB measurements are restricted to the low redshift range, $z$<~2, with the vast majority $z$<~1, while the CMB measurements of *Planck* and WMAP satellites correspond to a single very high redshift, $z$~1100, the location of the point of last CMB scattering. Then it will be difficult to distinguish between our two models, as the critical wide intermediate range, ~1.5<$z$<1100, is sparsely measured.

### 3.1 Usual CPL parameterisation of *w(a)*.

Present instrument resolution limits us to assume a simple shape for the *w(a)* timeline with a description using a minimum number of parameters. Combined datasets are integrated over a wide range of redshifts to find those optimum parameter values. It is usual to denote the present equation of state parameter by $w_o$, with the value at much earlier times denoted by $w_o+w_a$. Most astrophysical datasets, including recent *Planck* data [1-3], were analysed to deduce cosmological parameters using the simple two parameter 'CPL' form of parameterisation [101] given in equation (4).

$$w(a) = w_o + (1-a)\, w_a \qquad (4)$$

This parameterisation assumes a smooth, continuous variation of w(a) from $w_o+w_a$ at very early times, $a$<<1, through to $w_o$ today, with a midpoint value of $w_o+0.5w_a$ at $z$=1.

The 2013 and 2015 *Planck* data releases [1-3] include several dataset combinations where *Planck* data have been combined with other types of measurement and analysed using the CPL parameterisation. Those other datasets include: baryon acoustic oscillations (BAO); supernovae SNe1a studies such as the Supernova Legacy Survey, SNLS, and 580 supernova compiled in Union2.1; Galaxy clustering; the Hubble Space Telescope (HST); and WMAP CMB. Marginalised posterior distributions shown by the 68% (2σ) and 95% (1σ) likelihood contours for some of these data combinations are re-plotted in Fig. 3. in $w_o$-$w_a$ space.

Fig.3, upper plot shows *Planck* 2013 data release (from Fig.36 of [1]) while the lower plot shows the *Planck* 2015 data release (from Fig. 28 of [2]). The *Planck* 2015 data [2,3] includes several improvements in data treatment and calibration over the 2013 data [1]. Besides the cosmological constant ($w_o$=-1,$w_a$=0), we include on Fig. 3. upper plot the quintessence regime, +1≥ w ≥-1 , the phantom energy regime, w <-1, and, for completeness, the quintom regimes where, at some point in time, w crossed the value w=-1. We have also shown the HDIE predicted value from this work (§2.2) and from the earlier HDIE publication [29].

In the upper plot we have also plotted the 68% (solid black) and 95% (dashed black) likelihoods common to all three *Planck* 2013 dataset combinations. We note that both HDIE predicted values lie inside, or very close to, the common 68% likelihood 2σ region, while the cosmological constant lies just outside even the common 95% likelihood 1σ region. In the recent *Planck* 2015 data release, lower plot, the three dataset combinations show a much lower extent of overlap. One combination is fully consistent with a cosmological constant while the other two combinations are more consistent with phantom energy. Both HDIE predicted values lie approximately around the average of all three dataset combinations.

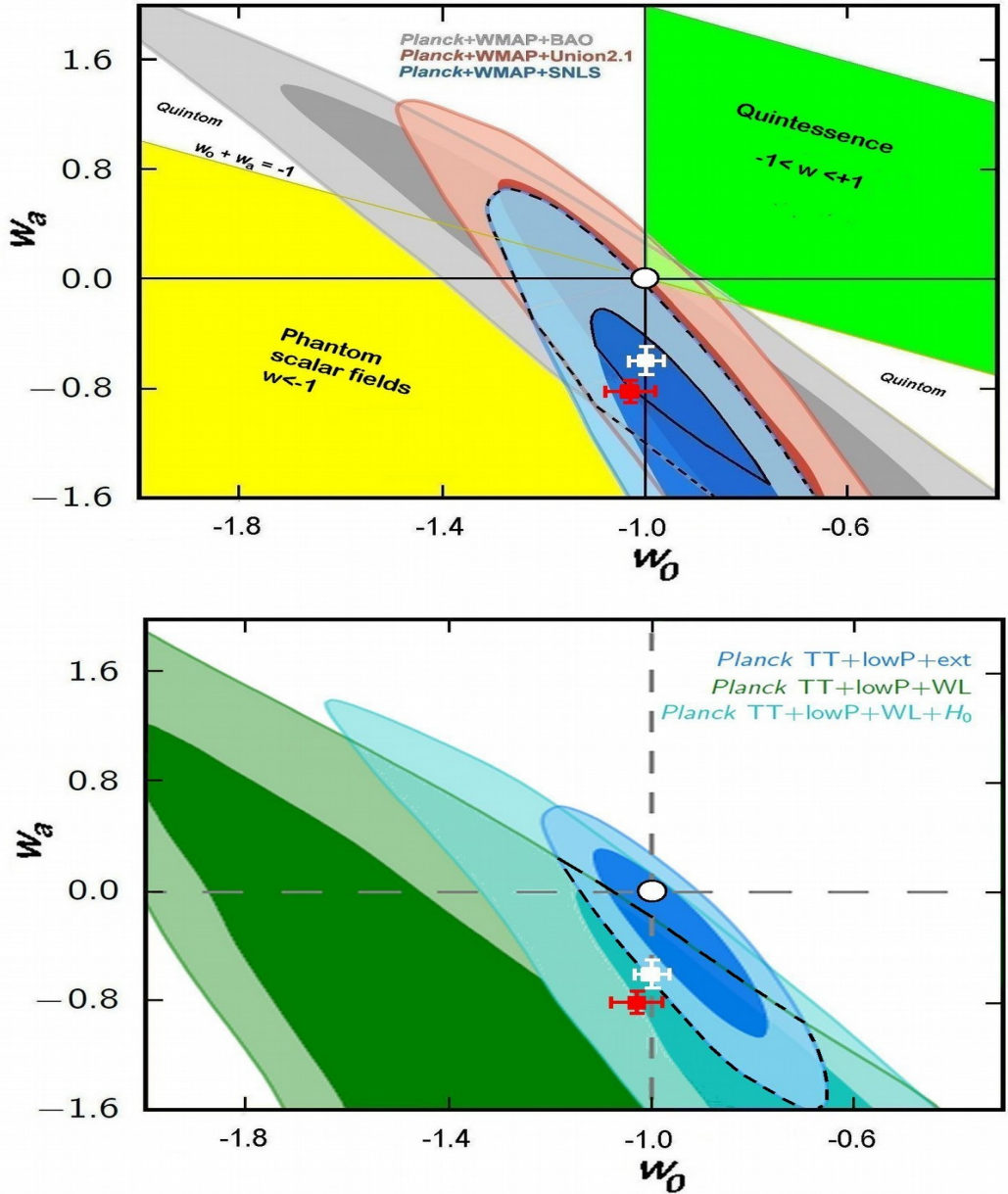

**Fig.3. HDIE predictions compared with *Planck* dataset combination results.** 2D marginalised posterior distributions are shown by the 68% and 95% likelihood contours as a function of $w_o$ and $w_a$ for different dataset combinations. Symbols: white circle, cosmological constant; white square, previous HDIE work [29]; red square, this HDIE work. *Upper plot:* 2013 results for the three data combinations: *Planck+WMAP+BAO*; *Planck+WMAP+Union2.1*; and *Planck+ WMAP+SNLS* from [1]. The areas bounded by the black dashed line and the black continuous line correspond to the 95% and 68% likelihoods, respectively, that are common to all three dataset combinations. *Lower plot:* 2015 results for the three data combinations: *Planck* TT + lowP +ext; *Planck* TT + lowP + WL; and *Planck* TT + lowP + WL + $H_0$ from [2]. Dashed line 95% common likelihood. Refer to *Planck* results [1, 2] for a full description of these data combinations.

Both groups of dataset plots in Fig.3. have a strong tendency for $w_a \leq 0$, favouring either a cosmological constant or phantom energy explanations over quintessence. Conservatively we can say that HDIE predicted values fit both 2013 and 2015 releases of *Planck* combined datasets at least as well as, and a little better than, ΛCDM, but it is clear that, as yet, these data using the CPL parameterisation are not significant enough to decide between ΛCDM and HDIE/baryon models.

**3.2. Mounting (weak) evidence for phantom Dark Energy.**

Here we list measurements that seem to favour a phantom DE. It is important to preface this list by emphasizing that, despite the clear trend for $w_a \leq 0$ in Fig.3, and the results listed here, no measurement yet exists that can exclude the cosmological constant to a level greater than 3σ. Note most of the listed items below are directly compatible with the HDIE/baryon predicted $w(a)$.

1) We start by noting that *Planck* CMB data is not strongly constrained without combining with other data sets. On its own this *Planck* data yields $w$=-1.54, +0.62/-0.50 corresponding to a ~2σ shift into the phantom regime [2], effectively averaging over the whole range 0<$z$<1100.

2) Some measurements of $w$ may be considered to be inconsistent with a cosmological constant. One survey of 146 Type Ia Supernovae over the range 0.03<$z$<0.65 when combined with BAO, *Planck* and $H_0$ finds $w$=-1.166+0.072/-0.069, inconsistent with $w$=-1 at 2.3σ level [102].

3) A wider combination of current and mature DE measurements finds that the value of $w$ strongly depends on the value adopted for the Hubble constant, $H_o$. That analysis concludes at ≥2σ confidence: either datasets still contain unaccounted for systematic errors, or $H_o$ < 71km/s/Mpc, or dark energy is phantom, $w$<-1 [103].

4) Another analysis of similar datasets also implies DE is phantom at 2σ level: $w$=-1.15±0.07 with the dataset combination *Planck* + WMAP + BAO + Union2.1 + HST and $w$=-1.16±0.06 with the dataset combination *Planck* + WMAP + BAO + SNLS + HST. [104]

5) A comparison between $H_0$ derived from *Planck* with that from more direct measurements favours $w$<-1 at the 2σ level, and, if DE is dynamic, finds 68% confidence level constraints of $w_o$ = -0.81 ±0:19 and $w_a$ = -1.9 ±1.1 [105] using the CPL parameterisation.

6) Recent BOSS measurements of quasars in the Lyman α forest [106] provide further support for $w_a$<0 with values in tension with the standard ΛCDM model, and a DE energy density at $z$=2.4 that is less than the energy density at $z$=0.

7) The most accurate NASA Hubble measurement of $H_0$ to date [107] is ~8% higher than the value derived from CMB measurement combinations of the *Planck* consortium [2]. This higher value is also supported by recently released data from the ESA GAIA spacecraft. One likely explanation for this 3.0σ difference is that it is caused by a dynamic and phantom DE.

8) A recent dynamic dark energy analysis for $w(a)$ has compared ten combinations of different types of measurement using the latest available datasets [108]. All ten dataset combinations average around $w$~-1 for $z$<1, but, at higher redshifts, all ten combinations exhibit clear phantom DE, $w$< -1.

**3.3. Proposed three parameter description of *w(a)*.**

It has been argued [109] that the usual two parameter CPL parameterisation, by virtue of its simplicity, makes a minimum assumption about the shape of $w(a)$, and, accordingly, has the advantage of being effectively neutral and not biasing the analysis towards any one particular explanation of DE. For these reasons CPL has been widely adopted and has enabled us to make comparisons between widely different measurement techniques and their combined datasets. A number of publications have suggested different parameterisations, often justified on theoretical grounds [110-116], while a case is made below for a simple three parameter parameterisation driven solely by the empirical evidence of the star formation data of Fig. 1.

We note that the BOSS measurements [117] exclude any significant phantom behaviour at low $z$, clearly restricting DE explanations towards a cosmological constant type behaviour below $z$~0.7, with limits to -0.97>$w$>-1.11 over the range 0.2<$z$<0.7. On the other hand most of the combined *Planck* dataset combinations (Fig.3.), although also compatible with the cosmological constant, allow for significant phantom DE. This suggests that, while there is a cosmological constant type of behaviour out to at least $z$=0.7, there could be a possible phantom behaviour at higher $z$ values, based on the list of §3.2.

This tension suggests there is a relatively rapid change at some intermediate $z$ value. Adopting a description that permits a sharper change than provided by CPL necessarily requires an increase in the number of parameters to more than two, and ideally described by four parameters as proposed previously [116] and illustrated by equation (5).

$$w(a) = w_o+(w_a/(1+\exp((a-a_t)/a_w))) \qquad (5)$$

This description maintains the limits of $w_o$ today and $w_o + w_a$ at much earlier times as in CPL. In comparison to the slow continuous change provided by CPL, the above four parameter form allows for a much sharper transition between the two values with the transition centered around $a_t$ and with the width of transition region set by $a_w$.

Unfortunately, the use of four parameters significantly complicates dataset merging and dataset comparisons. In order to simplify data analysis, we then further assume that the transition is so sharp, $a_w \rightarrow 0$, that we can ignore any variation of the data within that transition region. We justify taking this approach with the power law fits (red lines) to the observed star formation data of Fig.1. Now the description effectively resorts to an instantaneous transition at $a_t$, allowing for the simpler three parameter description of equation (6).

$$w(a)= w_o+w_a,\ a<a_t\ ;\quad w(a)= w_o,\ a>a_t. \qquad (6)$$

To illustrate the advantage of using this sharper transition over CPL as a means of distinguishing between ΛCDM and HDIE/baryon models, we now consider four specific parameter combinations:

1) Λ: $w_o$=-1, $w_a$=0. These parameter values correspond to the cosmological constant, (black lines Figs 1 & 4), closely fitting BOSS data and HDIE at low redshifts, $z$<1.35, but making a bad fit to HDIE at higher redshifts. Note that the zero value of $w_a$ makes this case independent of whether considering CPL or the three parameter description (and independent of $a_t$ ).

2) CPL $w_o$=-1, $w_a$=-0.45. Assuming $w_o$=~-1, then $w_a$=-0.45 is the most negative value of $w_a$ for CPL that still fits BOSS low $z$ data illustrated by the yellow wedge in Fig.4. This description (dashed grey line in Figs. 1 & 4) makes a very poor fit to HDIE at higher redshifts, $z$>~2.

3) CPL $w_o$=-1, $w_a$=-0.82. Best CPL approximation to HDIE predicted by §2.2 of this work (continuous grey line in Figs. 1 & 4). While closely fitting HDIE at very high redshifts, ($w_o+w_a$), and at very low redshifts ($w_o$), this parameterisation makes a poor fit over the important intermediate redshift range: ~2> z >~0.5. Note that this description also does not fit well with BOSS measurements as it lies outside the majority of BOSS measurements (yellow wedge in Fig.4.)

4) Three parameter sharp transition. $w=w_o=-1.03$ at $z<1.35$ ($a_t>0.43$), and $w=w_o+w_a$ at $z>1.35$ ($a_t<0.43$), where $w_a=-0.79$. These are the specific values that closely describe HDIE values at all redshifts, as predicted from Fig.1. star formation history (red lines in Figs. 1 & 4). This case also makes a good fit to BOSS measurements, falling completely inside the yellow wedge of Fig.4.

Star formation histories numbered 1-4 plotted in Fig. 1. would cause HDIE to have varied as described by these four parameterisations. In Fig.4, upper plot, we plot energy density contributions and energy totals for these parameterisations, all with present values of 68.3% DE and 31.7% all mass (i.e. independent of whether DM is present). For each of these descriptions we see that after matter energy density is included, total energy densities are both very similar to each other and to a cosmological constant (black line).

In order to better identify measurable differences, Fig.4, lower plot, shows the Hubble parameter values of descriptions 2-4, relative to that expected for a cosmological constant. The polynomial fit to data of Fig.1. is included on the figures (blue line Figs. 1 & 4) for comparison. Recent Baryon Oscillation Spectroscopic Survey (BOSS) measurements of clustering of galaxies [117] are fully consistent with a cosmological constant type of behaviour over the redshift range $0.2 < z < 0.7$ to an accuracy of 1%, illustrated by the yellow wedge area in Fig. 4, upper plot. Also plotted on the lower plot are the expected threshold resolutions of three next generation instruments: Euclid [118]; WFIRST [119]; BigBOSS [120].

We summarise our observations of Figs 1, 3 and 4 in Table 2. Case 2 lies midway between the cosmological constant and the HDIE points in Fig. 3. and all four cases make a reasonable fit to the *Planck* data combinations. Then the three parameter sharp transition, case 4, is the only one to make a good fit to both HDIE data and BOSS measurements. Although this three parameter description provides the best fitting, Fig.4, lower plot, shows that it will be very difficult to distinguish between the two models. The expected difference in Hubble parameter between case 4 and the ΛCDM model lies close to, or below, the resolutions limits of the next generation of DE instruments. In contrast, cases 2 and 3 would be easily distinguished from ΛCDM by measurement but neither makes both a fair representation of the HDIE predicted $w(a)$ variation and also fits BOSS data.

| Case: | Fit to HDIE | Fit to BOSS | Fit to Planck | Difference in $H(a)$ relative to Λ |
|---|---|---|---|---|
| 1) Λ: $w_o=-1$, $w_a=0$ CPL or 3 parameter | bad | good | reasonable | -- |
| 2) CPL $w_o=-1$, $w_a=-0.45$ | bad | marginal | reasonable | easily measured |
| 3) CPL $w_o=-1$, $w_a=-0.82$ | poor | bad | reasonable | easily measured |
| 4) 3 parameter fit to HDIE, $w_o=-1.03$, $w_a=-0.79$, $a_t=0.43$ | good | good | reasonable | ~ at resolution limit of next missions |

**Table.2.** The four cases compared for data fit and ease of measurement.

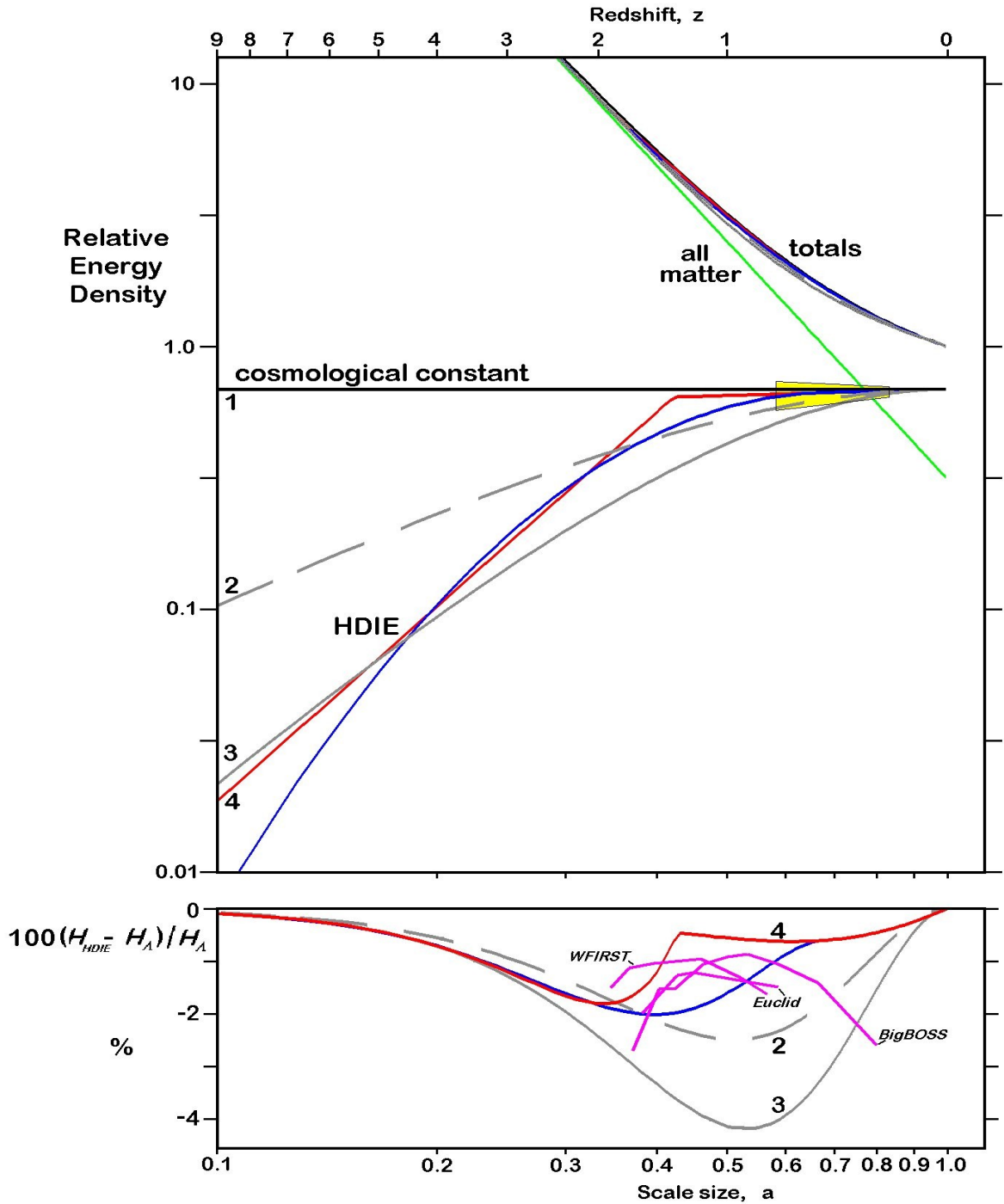

**Fig 4.** Comparison of parameterisations: energy density contributions and Hubble parameter.

*Upper plot.* Energy densities relative to total today(=1) versus scale size for all matter (green), cosmological constant, case1(black line), and HDIE using fits to stellar mass density data of Fig.1: red line, case 4, power laws, $w=-1.03$, $z<1.35$, and $w=-1.82$ $z>1.35$; blue line, simple polynomial fit; and grey lines, CPL type behaviour, $w=w_o+(1-a)w_a$, with continuous grey line $w_o=-1.0$, $w_a=-0.82$, case 3, and dashed grey line $w_o=-1.0$, $w_a=-0.45$, case 2. Total energy densities are also shown for all matter plus each DE variation. Yellow shaded area corresponds to the DE limit $-0.97>w>-1.11$ set by BOSS measurements [117] over the redshift range: $0.2<z<0.7$.

*Lower plot.* Corresponding variations in Hubble parameter $H(a)$ for the HDIE fits, cases 2,3,4, plotted as a percentage difference from that expected for a cosmological constant, and compared with the detection resolutions (purple lines) for Euclid[118], WFIRST[119], and BigBOSS[120].

# 4. Discussion

The work reported here employs two foundational principles of information theory: Landauer's Principle and the Holographic Principle. Landauer showed that information is physical[31,32]. The ultimate significance of all physical quantities is determined by elementary questions with simple binary yes/no answers. This approach is epitomised by Wheeler's slogan "It from Bit", implying information may even be the more fundamental language in which to express physics [121]. If the HDIE/baryon model is found to be the correct model, it will also provide strong support for the holographic principle. Otherwise, experimental proof of the Holographic Principle requires making near impossible noise/granulation measurements down to, or below, the Planck length ($10^{-35}$m).

We have found that information energy could account for a major part of the universe (DE+DM) but otherwise remains insignificant in our everyday lives. For example, the heat produced from information erasure within our computers is typically ~$10^{-11}$ of the overall electronic heat generation [28,29]. The low bit energy equivalence has lead to nearly half a century elapsing after Landauer's principle was proposed before it was finally experimentally verified [37-39].

In this work we found that star formation had a faster growth rate at earlier times, changing around $z\sim1.35$ to the slower rate of $a^{+1.08\pm0.16}$ to closely match the $a^{+1}$ rate required for a constant HDIE energy density, assuming a holographic universe, $N \propto a^2$. Is it just a coincidence that the present growth rate of stellar mass density is such as to provide near constant HDIE energy density, that entropy estimates enable HDIE to account for observed DE, and that the change in growth rate occurred just when DE initiated acceleration? Indeed, the most likely result of an HDIE explanation would be to cause feedback that caps the growth rate around the observed $a^{+1}$ value for constant HDIE density. In section 2.4 we saw the appeal of the HDIE/baryon model with many advantages over the ΛCDM model. Clearly, the key to verifying the HDIE/baryon model lies in future DE measurements at the higher redshifts where this model predicts a phantom DE. Then, if expansion measurements are eventually found to fit the specific form of dynamic $w(a)$ predicted by the HDIE/baryon model, HDIE must also be strong enough to make a significant contribution towards explaining many of the DM attributed effects. In comparison, the DM of the ΛCDM model has yet to be identified. Besides the null WIMP measurements [17-20] to date, now the energy range of any possible axions has also been significantly restricted [122].

The present disagreement (up to 3.7σ) between Hubble constant values obtained by different techniques, measuring over different red-shift ranges assuming ΛCDM [106,123,124], may be explained by an earlier phantom dark energy like HDIE. Besides the weak phantom evidence of §3.2 and combined *Planck* datasets of Fig.3., there are also other observations, of a more indirect nature, that are also difficult to reconcile with the ΛCDM model [125-127], including the latest measurements of many satellite galaxies found to be located in the same orbital plane around their parent galaxies [93,94].

CPL has served us well up to the present day, but here we are trying to resolve small model differences over a specific redshift range. The four cases considered in Fig.4. and Table 2. show that the continued use of the CPL parameterisation biases interpretation towards a cosmological constant, away from the possibility of an HDIE explanation, and contrary to the usual assumption of CPL neutrality.

The next generation of dark energy instruments include Euclid; WFIRST; BigBOSS; LSST[128]; Dark Energy Survey [129]; and the James Webb Space Telescope [130]. As the design, construction, data accumulating and processing operations of these instruments involves long timescales, accurate DE measured parameters will not be available for some time. We should therefore attempt a low cost re-analysis of some of the existing datasets using the proposed three parameter parameterisation. Even if this exercise fails to reduce the number of DE models, any resultant restriction in dark energy parameter range might contribute to optimizing the design and/or operation of those next generation instruments with subsequent enhanced scientific return.

Existing data analysis processes are already mature and well organised to deduce cosmological parameters, and it should be relatively easy to modify data analysis, changing from the usual CPL form to the three parameter form. At its simplest, analysis would involve repeating the present procedure a number of times, each with a different fixed value of the transition time, $a_t$, and thus each effectively still a two parameter $w_o$, $w_a$ analysis as done presently. DE instrument 'figures of merit' are defined as the inverse of the areas enclosed by the likelihood contours in $w_0$-$w_a$ space (as in Fig.3.). In the same way the optimum value of $a_t$ will be found to be the value that exhibits the minimum enclosed likelihood contour area in $w_o$-$w_a$ space.

It is interesting to note how the laws of thermodynamics apply to the universe as a whole. The first law appears to be violated on the scale of the universe since, in recent times, total DE (whatever the explanation) appears to increase with the expanding universe as $\sim a^3$. However, in general relativity DE may increase without seeming to conserve energy, because of the continual exchange of energy between matter and changing space-time or gravitational field [131]. The anti-de-Sitter/conformal field theory (Ads/CFT) correspondence translates a multidimensional space with gravity to another multidimensional space without gravity but with one less dimension [132]. This has led to the suggestion that, by combining the holographic principle with Landauer's principle, in a similar way to the entropic energy of HDIE, the information carried by space-time is equivalent to an entropic force, an entropic or emergent quantum gravity [133,134], which, like HDIE, possibly also accounts for DM attributed effects [134]. The work reported here only uses the simplest prediction from the Holographic principle, that information contained by a given volume is proportional to bounding area, $N \propto a^2$. In this way we have concentrated on measured data and avoided the complicated theoretical considerations of string theory, etc, inherent in the holographic approach. We have used the energy equivalence of large numbers of bits without actually performing a measurement on any bits. This has enabled us to discuss bits effectively as elementary systems without the need to distinguish between classical bits and quantum qu-bits [135].

On the other hand, the second law should apply universally. If algorithmic information is also governed by the second law, a simple Gedanken experiment has revealed a further connection between information and universe acceleration [28]. This thought experiment considered the algorithmic information describing the baryons in the whole universe, but also applies to any large co-moving volume. It was easily shown that the observed increasing star formation would have resulted in a decrease in the number of algorithmic information bits if the expansion had not started to accelerate. In order to ensure the 2nd law is satisfied with no decrease in algorithmic information bit number, an approximate extra doubling of universe volume is found to be required, as is indeed observed to have resulted directly from the recent period of acceleration.

In the ΛCDM model Λ is just a constant of equation (2) with no satisfactory physical explanation. In equation (3) the HDIE/baryon model replaces this term with a dynamic information dark energy. Despite this significantly different approach there does exist an interesting similarity between Λ and information energy. The characteristic energy of Λ is given by $\Delta_\Lambda = (15 \rho_{tot} \hbar^3 c^5/\pi^2)^{1/4}$. This has a present value of ~$3 \times 10^{-3}$eV, considered too small to relate to any relevant particle physics [136]. Equation (2) effectively ignores star formation and without such structure formation we could assign to matter a representative temperature, $T_r$~35, the temperature that provides a radiation energy density equal to that due to the universe matter density, $\rho_{tot}$. [ given by $\rho_{tot} c^2 = \alpha T_r^4$ where the radiation constant $\alpha=4\sigma/c$]. When we describe the Stefan-Boltzmann constant, $\sigma$, in terms of fundamental constants, we obtain an information bit energy of $k_B T_r ln2 = (15 \rho_{tot} \hbar^3 c^5/\pi^2)^{1/4} ln2$, identical in definition and value to $\Delta_\Lambda$, but for the addition of the *ln2* term[137,138]. While this simple similarity ignores star formation, the work reported here concentrates on the information associated with the much higher temperature stellar heated gas and dust.

Finally, it is usually assumed that, if DE energy density remains constant into the future, the resulting continuing acceleration will eventually cause the so called 'big rip' of the universe. In the case of our HDIE explanation, the fraction of baryons in stars will continue to increase as $a^{+1}$ for the immediate future, providing constant HDIE energy density. But that fraction, by definition, cannot exceed unity. At later times we expect a falling off in star formation and total star numbers, and therefore leaving us eventually with expansion without acceleration, perhaps more analogous to a 'slow tear' than a 'big rip'.

## 5. Conclusion.

Given our current lack of understanding of DM and DE phenomena, there is a clear case for looking beyond ΛCDM and pursuing alternatives. Rather than theory driven, the HDIE/baryon model is primarily data driven and provides a common explanation for many effects previously attributed separately to DM and DE. This model has several advantages, not least in simplicity of concept, and naturalness, or reliance on mostly proven physics. We have used the ratio of HDIE to baryons to account for acceleration due to DE and to provide effective mass fractions to account for DM-like effects. HDIE fits *Planck* data combinations' $w_o$-$w_a$ plots at least as well as Λ, and the prediction of earlier phantom DE is consistent with some other measurements.

The usual CPL parameterisation clearly biases data interpretation towards the standard ΛCDM model while the proposed three parameter ($w_o$, $w_a$, $a_t$) parameterisation should be neutral, at least between ΛCDM and HDIE/baryon models, and eventually provide a clear distinction between them. Although the immediate next generation of instruments might still not possess sufficient resolution for this model separation, future higher resolution measurements at intermediate redshifts will identify the more appropriate model. If dark energy is eventually shown to have been phantom at earlier times, and, if there has still been no confirmed detection of dark matter, then the HDIE/baryon model will provide an information based explanation joining together both DE and DM aspects of the Dark Side.

**Acknowledgments:** The author is grateful for an Emeritus Professorship from the University of Sussex.